\documentclass[a4paper,10pt]{article}
\usepackage[utf8]{inputenc}
\usepackage{amsmath,amsthm}
\usepackage{authblk} \usepackage{blindtext}
\usepackage[a4paper,left=3cm,right=3cm,top=3cm,bottom=3cm]{geometry}

\usepackage{graphics}
\clearpage{}\usepackage{hyperref}
\usepackage{doi}
\usepackage{cite}
\usepackage{stmaryrd} \usepackage{dsfont} \usepackage{graphicx}
\usepackage{enumerate}
\usepackage{subfigure}

\usepackage{mdframed}

\usepackage{chngcntr}
\usepackage{apptools}

\DeclareMathOperator{\ALukyanov}{A_\text{CSM}}
\DeclareMathOperator{\thetastar}{\theta^\ast}

\newcommand{\inproduct}[2]{\left\langle #1 , #2 \right\rangle}

\newcommand{\divergence}[1]{\nabla \cdot #1}

\newcommand{\transpose}{{\sf T}}

\newcommand{\RR}{\mathds{R}}    \DeclareMathOperator{\ca}{Ca} \newcommand{\visc}{\eta}

\DeclareMathOperator{\thetaeq}{\theta_{eq}}

\newcommand{\vsigma}{v_\Sigma}   

\clearpage{}

\title{Boundary conditions for dynamic wetting -\\A mathematical analysis}
\author[1]{Mathis Fricke\thanks{fricke@mma.tu-darmstadt.de}}
\author[1]{Dieter Bothe\thanks{bothe@mma.tu-darmstadt.de}}

\affil[1]{Department of Mathematics, TU Darmstadt, Germany}
\date{}

\begin{document}

\maketitle

\begin{abstract}
The moving contact line paradox discussed in the famous paper by Huh and Scriven has lead to an extensive scientific discussion about singularities in continuum mechanical models of dynamic wetting in the framework of the two-phase Navier-Stokes equations. Since the no-slip condition introduces a non-integrable and therefore unphysical singularity into the model, various models to relax the singularity have been proposed. Many of the relaxation mechanisms still retain a weak (integrable) singularity, while other approaches look for completely regular solutions with finite curvature and pressure at the moving contact line. In particular, the model introduced recently in (Lukyanov, Pryer, Langmuir 2017) aims for regular solutions through modified boundary conditions. \\
The present work applies the mathematical tool of compatibility analysis to continuum models of dynamic wetting. The basic idea is that the boundary conditions have to be compatible at the contact line in order to allow for regular solutions. Remarkably, the method allows to compute explicit expressions for the pressure and the curvature locally at the moving contact line for regular solutions to the model of Lukyanov and Pryer. It is found that solutions may still be singular for the latter model.

 \end{abstract}

This preprint was accepted for publication in \emph{The European Physical Journal Special Topics}.\newline When citing this work, please refer to the journal article: \textbf{DOI:} \href{http://dx.doi.org/10.1140/epjst/e2020-900249-7}{10.1140/epjst/e2020-900249-7}.

\section{Introduction}
Starting with the work by Huh and Scriven \cite{Huh.1971}, the scientific discussion about singularities became central for the continuum mechanical modeling of dynamic wetting (see, e.g., \cite{Gennes.1985,Shikhmurzaev.2006,Shikhmurzaev.2008,Bonn.2009,Velarde.2011,Snoeijer.2013,Eggers2015}). While it is generally accepted that the non-integrable singularity in the viscous dissipation introduced by the no-slip condition (see \cite{Huh.1971}) is unphysical for a viscous fluid, there are different approaches to relax the singularity. See \cite{Shikhmurzaev.2008,Bonn.2009,Velarde.2011,Snoeijer.2013,Sui.2014,Eggers2015} and references therein for an overview of existing models and the field of dynamic wetting in general. In particular, the articles and controversial discussion notes contained in \cite{Velarde.2011} provide a comprehensive overview about methods and open questions. For the sake of brevity, we only consider two particular approaches in this paper. Note, however, that the method of compatibility analysis is general in nature and is applicable to other modeling approaches as well.\\
\\
One of the most prominent choices is the Navier slip law which allows for tangential slip at the solid boundary according to
\begin{align}
\inproduct{v}{n} &= 0,\label{eqn:comp_impermeability}\\
-\lambda (v-w)_\parallel &= 2 \eta (Dn)_\parallel\label{eqn:comp_navier_slip},
\end{align}
  where $\lambda > 0$ is a friction coefficient, $w$ is the velocity of the solid boundary, $n$ is the unit outer normal and $D = \frac{1}{2}(\nabla v + \nabla v^\transpose)$ is the rate-of-deformation tensor. The parameter $L=\eta/\lambda$ is called slip-length and controls the amount of tangential slip at a given shear rate.\\
\\
Asymptotic methods allow to obtain information about the local structure of the solution near the contact line for the no-slip and Navier slip model (see \cite{Moffatt.1964,Cox.1986}). For quasi-stationary solutions, it has been shown by asymptotic methods  \cite{Huh1977} that a finite and positive slip length leads to an integrable singularity with the pressure behaving like
\begin{align}
p \propto \log r,
\end{align}
where $r$ is the distance to the contact line. In this case, the force balance at the interface implies that the curvature is also infinite at the contact line to oppose the singular force due to pressure. Note that this irregular behavior of the solution might lead to problems when the model is solved numerically \cite{Sprittles.2011}. While some authors argue that this ``weak'' type of singularity has little influence on the macroscopic flow \cite{DussanV.1976}, there is also a large body of research looking for continuum mechanical models which do not even show weak singularities (see, e.g., \cite{Shikhmurzaev.2006,Velarde.2011,Rednikov2013,Lukyanov.2017}).\\
\\
One approach chosen to regularize the moving contact line problem is to separately balance the mass contained in an interfacial layer. A framework for this kind of modeling is provided by the non-equilibrium thermodynamics of surfaces \cite{Bedeaux.1986}. Mathematically, the mass within the interfacial layer is expressed as a density per unit \emph{area} associated with a sharp interface in the continuum limit. The resulting model, known as the Interface Formation Model (IFM) \cite{Shikhmurzaev.1993,Shikhmurzaev.2008}, adds another level of complexity to the description since it requires to solve additional balance equations on moving interfaces. Physically, the idea is that the process of formation or disappearance of a piece of interface has a relaxation timescale which leads to dynamic surface tensions. Therefore, the model predicts a dynamic contact angle which is governed by a dynamic version of the Youngs equation, i.e.\
\[ \sigma_1 \cos \theta_d = \sigma_3 - \sigma_2, \]
where the surface tension coefficients $\sigma_i$ for the respective surface layers depend on the local state of the interface. Notably, the pressure and curvature at the moving contact line is claimed to be regular (see \cite{Shikhmurzaev.2008} for a detailed discussion).\\
\\
In the present article, we study a recently proposed model by Lukyanov and Pryer \cite{Lukyanov.2017}, which can be understood as a quasi-stationary adaptation/simplification of the full IFM. The basic idea is to allow for non-zero interfacial mass densities but to require that these are \emph{constant} in space and time. This assumption allows to substantially simplify the governing equations of the IFM. In particular, there is no need to solve the mass transport equation on the surfaces. Instead, the velocity associated with transport along the surfaces can be eliminated resulting in a modified set of boundary conditions for the stationary Stokes equations. Note that the fundamental difference to a model without interfacial mass is that the impermeability condition \eqref{eqn:comp_impermeability} at the solid boundary is relaxed because mass can be transported from the bulk to the surface phase. The model is introduced and applied to nanodroplets in \cite{Lukyanov.2017}. Remarkably, it is stated that there ```\textit{a small, albeit natural, change in the boundary conditions is all that is necessary to completely regularize the problem}"\cite{Lukyanov.2017}. However, the authors neither prove the latter claim nor provide a numerical convergence study for the pressure and the curvature at the contact line. Before we discuss the latter model in more detail, the method of compatibility analysis is introduced and applied to the ``standard Navier slip model''.

\paragraph{Compatibility conditions for partial differential equations:} So-called \emph{compatibility conditions} appear naturally in the study of initial-boundary value problems for partial differential equations (PDEs) if one requires higher regularity of a solution, see \cite{Temam2006} and the references given there for an introduction to the topic. A simple example is the one-dimensional convection equation: 
\begin{equation}
\label{eqn:comp_advection_example}
\begin{aligned}
\frac{\partial u}{\partial t} + \frac{\partial u}{\partial x} = f, \quad 0 < x < 1, \, t > 0,\\
u(0,t) = 0, \quad u(x,0) = u_0(x).
\end{aligned}
\end{equation}
Here, we briefly recall the discussion given in \cite{Temam2006}. A first compatibility condition is obtained from taking the limit of the initial and boundary condition as $x \rightarrow 0$ and $t \rightarrow 0$, respectively.  If $u$ is continuous up to $(0,0)$ it follows that
\begin{align}
\label{eqn:comp_example_condition_1}
u_0(0) = 0.
\end{align}
Moreover, if $u$ is $\mathcal{C}^1$ up to $(0,0)$ we have $(\partial_t u)(0,t) = 0$ (since $u(0,t) = 0)$ and hence
\[ \frac{\partial u}{\partial x} (0,t) = f(0,t), \quad t > 0 \]
leading (as $t \rightarrow 0$) to the second compatibility condition
\begin{align}
\label{eqn:comp_example_condition_2}
u'_0(0) = f(0,0).
\end{align}
One can show that \eqref{eqn:comp_advection_example} is well-posed for smooth data $f$ and $u_0$ \cite{Temam2006}. Moreover, the compatibility conditions \eqref{eqn:comp_example_condition_1} and \eqref{eqn:comp_example_condition_2} are necessary and sufficient for $u$ to be $\mathcal{C}^1$; see \cite{Temam2006}. For example, in case $f=1$ the solution\footnote{The solution of the first-order PDE \eqref{eqn:comp_advection_example} can be found using the method of characteristics (see, e.g, \cite{evans10}).} of \eqref{eqn:comp_advection_example} reads as
\[ u(x,t) = \begin{cases} u_0(x-t) + t & \text{if} \quad x > t, \\ x & \text{if} \quad t > x.  \end{cases} \]
Hence the conditions \eqref{eqn:comp_example_condition_1} and \eqref{eqn:comp_example_condition_2} ensure continuity and continuous differentiability of the solution, respectively.
 
\section{Compatibility analysis for the standard slip model}
\label{sec:standard_model}
\begin{figure}[hb]
\centering
\includegraphics[width=6cm]{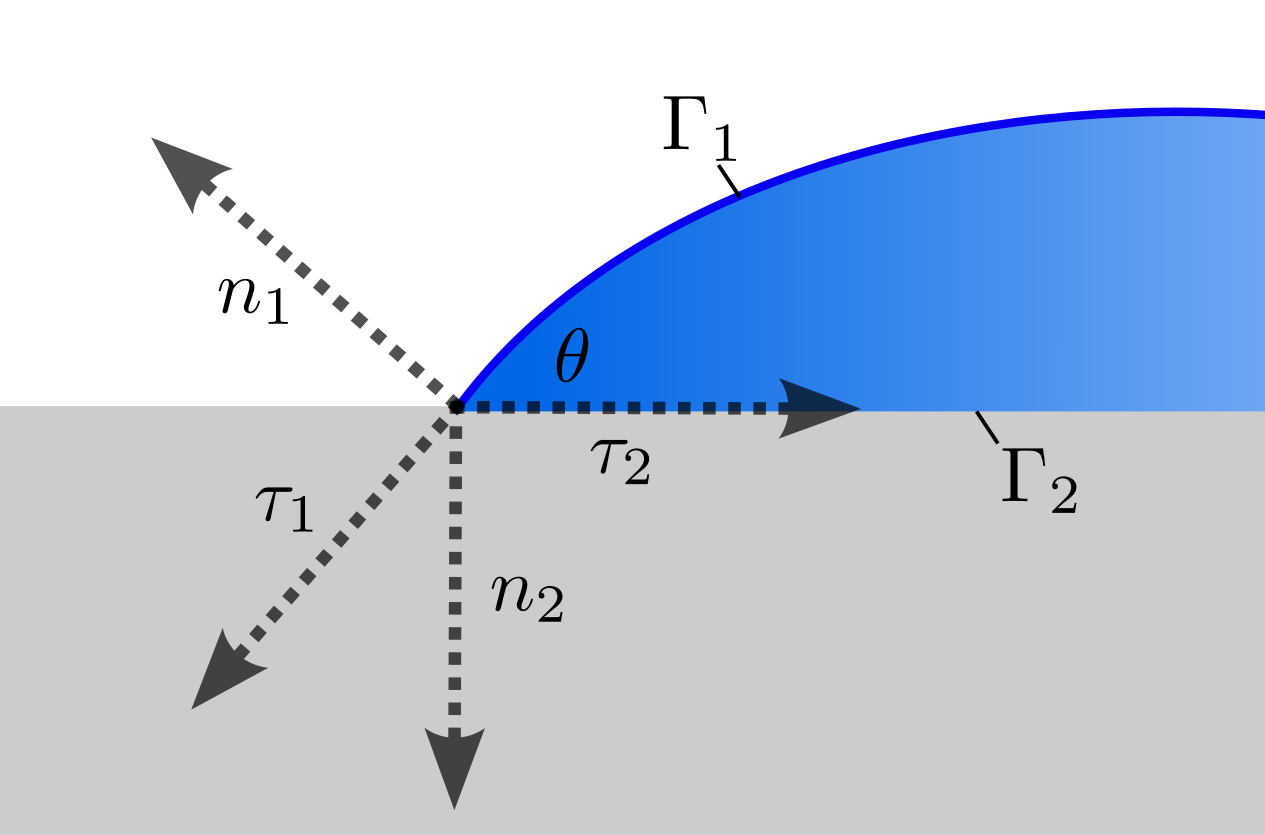}
\caption{Notation.}
 \label{fig:notation}
\end{figure}

For the following calculations, we consider the geometrical configuration depicted in Figure \ref{fig:notation}, where for simplicity we only consider the case of two spatial dimensions. We choose Cartesian coordinates in a reference frame co-moving with the contact line, i.e.\ the solid boundary is moving with velocity $(V_0,0)$ to the right, where $V_0$ is the speed of the contact line relative to the wall. Moreover, the interface normal and tangential vectors at the contact line have the form
\begin{align*} 
n_1(0,0)^\transpose  = (-\sin \theta, \cos \theta), \ \tau_1(0,0)^\transpose  = - (\cos \theta,\sin \theta),\\ 
n_2(0,0)^\transpose =(0,-1), \ \tau_2(0,0)^\transpose  = (1,0),
\end{align*} 
where $\theta$ is the contact angle. To simplify the analysis, we assume that the solid boundary is flat, i.e.\ $\kappa_2 = - \nabla_{\Gamma_2} \cdot n_2 = 0$.

\paragraph{Compatibility analysis applied to dynamic wetting:} The basic idea of the compatibility analysis applied to the wetting problem is to consider a local linear expansion of the (divergence free) velocity field \emph{at} the contact line, i.e.\
\begin{align}
\label{eqn:comp_linear_field}
\left(v_1,v_2\right)^\transpose = V_0 \left(c_1+c_2 \frac{x}{L} + c_3 \frac{y}{L}, c_4 + c_5 \frac{x}{L} - c_2\frac{y}{L} \right)^\transpose  + o(|x|+|y|),
\end{align}
where $V_0 \neq 0$ is the contact line speed and $L>0$ is the slip length. Therefore, the unknown coefficients $c_i$ are dimensionless. The crucial property is:\\
\\
If the solution is \emph{regular}, i.e.\ continuously differentiable up to the contact line, it allows for a local expansion of type \eqref{eqn:comp_linear_field} and obeys all boundary conditions at the contact line, where the free surface and the solid boundary meet (see \cite{Fricke.2019}).\\
\\
This requirement leads to a system of equations for the unknown coefficients $c_i$. We emphasize that the solvability of the latter system of equations is a \emph{necessary} condition for the existence of regular solutions (in the sense defined above). To prove the existence of regular solutions is a separate task that is not addressed in the present article. Note that the velocity itself and the velocity gradient at the contact line can be expressed as
\begin{align}
v(0,0) = V_0 (c_1,c_4)^\transpose, \quad \nabla v (0,0) = \frac{V_0}{L} \begin{pmatrix} c_2 & c_3 \\ c_5 & -c_2 \end{pmatrix}.
\end{align}
Clearly, higher-order terms in \eqref{eqn:comp_linear_field} do not contribute to $\nabla v(0,0)$. As we will see below, most of the boundary conditions only involve the symmetric part of $\nabla v$, the latter tensor being the rate-of-strain tensor $D$. At the contact line, it can be written in terms of the coefficients according to
\[ D(0,0) = \frac12 (\nabla v + \nabla v^\transpose)(0,0) = \frac{V_0}{2L} \begin{pmatrix} 2c_2 & c_3+c_5 \\ c_3+c_5 & -2c_2 \end{pmatrix}. \]
Due to the structure of $D(0,0)$, it is convenient to formally introduce a new unknown
\[ c_{35} := c_3 + c_5 \]
and to state the problem in terms of the unknowns $\{c_1,c_2,c_{35},c_4 \}$. 

\paragraph{Application to the standard slip model in the free surface formulation:} We consider all boundary conditions evaluated at the contact line.
\begin{enumerate}[(i)]
 \item The kinematic boundary condition\footnote{Here $V_{\Gamma_1}$ denotes the normal speed of the free surface.} at the contact line implies (in a co-moving frame)
 \[ 0 = V_{\Gamma_1} =  \inproduct{v}{n_1} = - c_1 \sin \theta + c_4 \cos \theta. \]
  \item Due to the impermeability condition $v \cdot n_2 = 0$ on $\Gamma_2(t)$ (evaluated at the contact line) it follows that $c_4= 0$.
 \item The zero tangential stress condition, i.e.\
 \[ 0 = 2 \eta \inproduct{\tau_1}{D n_1} \quad \text{on} \quad \Gamma_1(t), \]
  applies at the free boundary in the \emph{absence} of surface tension gradients. Evaluating the latter condition at the contact line yields
 \begin{align}
  \label{eqn:comp_zero_stress_condition_coefficients}
 0 = 2 c_2 \sin 2\theta - c_{35} \cos 2\theta.
 \end{align}
 \item The Navier slip condition evaluated at the contact line reads as
 \begin{align}
  \label{eqn:comp_navier_slip_standard_model}
  V_0 - \inproduct{v}{\tau_2}(0,0) = 2 L \inproduct{\tau_2}{D n_2}(0,0) \quad \Leftrightarrow \quad c_1 - c_{35} = 1.  
 \end{align}
 \item The normal stress condition (for $p_{\text{ext}} = 0$), i.e.\
 \begin{align}
 -p + 2 \eta \inproduct{n_1}{D n_1} = \sigma \kappa_1 \quad \text{on} \quad \Gamma_1(t),
 \end{align}
 provides the link between the two unknowns $p$ and $\kappa$.
\end{enumerate}

Hence we obtain a system of $4$ linear equations for $4$ unknowns ($c_1,c_2,c_{35},c_4$). The linear system reads as
\begin{align}
\label{eqn:comp_linear_system_standard_model}
\begin{pmatrix}
-\sin \theta & 0 & 0 & \cos \theta  \\ 0 & -2 \sin 2\theta & \cos 2\theta & 0 \\ 0 & 0 & 0 & 1  \\ 1 & 0 & -1 & 0 \\ \end{pmatrix}
\begin{pmatrix}
c_1 \\ c_2 \\ c_{35} \\ c_4
\end{pmatrix}
= \begin{pmatrix} 0 \\ 0 \\ 0 \\ 1\end{pmatrix}.
\end{align}
The determinant of the system matrix is given by
\[ \det A_{\text{Slip}} = 2 \sin\theta \sin 2\theta.  \]
For $\theta \not\in \{0,\pi/2,\pi \}$ the solution reads as
\begin{align}
c_1 = c_4 = 0, \ c_{35} = -1, \ c_2 = - \frac{\cot 2\theta}{2}.
\end{align}
Note that the coefficient $c_2$ becomes singular for $\theta \rightarrow \pi/2$. It is easy to show that no solution of the linear system exists for $\theta = \pi/2$. The latter case is distinguished mathematically since the tangential stress and Navier slip conditions produce linearly dependent equations which are incompatible (note that $c_1$ = 0). For the limiting cases $\theta \in \{0,\pi\}$, it is straightforward to show that \eqref{eqn:comp_linear_system_standard_model} has a family of solutions given by
\[ (c_1,c_2,c_{35},c_4) = (1,\lambda,0,0), \quad \lambda \in \RR.  \]

\paragraph{Qualitative behavior of regular solutions:} The evaluation of the boundary conditions \emph{at} the contact line only provides information about $v(0,0)$ and $D(0,0)$. Since the impermeability condition 
\[ \inproduct{v}{n_2} = 0 \]
applies along the whole solid boundary $\Gamma_2(t)$, it follows by differentiation with respect to $\tau_2$ (since $\kappa_2 = 0$) that
\[ \inproduct{(\nabla v) \tau_2}{n_2} = 0 \quad \text{on} \quad \Gamma_2(t). \]
The latter condition evaluated at the contact line shows 
\[ c_5 = 0. \]
So we found the local linear expansion of the velocity field for a regular solution (if existent) to the standard slip model. It has been shown in \cite{Fricke.2019}, that the rate-of-change of the contact angle can be computed from $\nabla v$ at the contact line according to
\begin{align}
\label{eqn:comp_theta_dot}
\dot\theta = -\inproduct{(\nabla v) \tau_1}{n_1} = \frac{V_0}{L} \left(- c_2 \sin 2\theta - c_3 \sin^2 \theta  + c_5 \cos^2 \theta \right).
\end{align}
Application of \eqref{eqn:comp_theta_dot} to the solution derived above yields (for $\theta \notin \{0,\pi/2,\pi \}$)
\begin{align}
\label{eqn:comp_theta_dot_slip_model}
\dot\theta = \frac{V_0}{2L}(\cos 2\theta+2 \sin^2 \theta) = \frac{V_0}{2L}.
\end{align}
The relation \eqref{eqn:comp_theta_dot_slip_model} shows that a (nontrivial) quasi-stationary solution of the model cannot be regular since $V_0 \neq 0$ implies $\dot\theta \neq 0$. Moreover, relation \eqref{eqn:comp_theta_dot_slip_model} is unphysical since the contact angle cannot relax to equilibrium if the thermodynamic condition for the contact line velocity, given by (see \cite{Ren.2007,Shikhmurzaev.2008,Ren.2010,Ren2011,Fricke.2019} for a discussion of entropy production at the contact line)
\begin{align}
V_0 \geq 0 \quad \text{for} \quad \theta \geq \thetaeq, \quad V_0 \leq 0 \quad \text{for} \quad \theta \leq \thetaeq,
\end{align}
is satisfied. This result shows that regular solutions to the standard model are unphysical and a singularity must be present at the contact line even if the contact angle is variable (see \cite{Fricke.2019} for a detailed discussion).

\paragraph{Effect of surface tension gradients:} To conclude the discussion of the standard Navier slip model, we consider the case of surface tension gradients at the free surface. In this case, the tangential stress condition reads as
 \[ \frac{\partial \sigma}{\partial \tau_1} = 2 \eta \inproduct{\tau_1}{D n_1} \quad \text{on} \quad \Gamma_1(t). \]
Consequently, the right-hand side of equation \eqref{eqn:comp_linear_system_standard_model} is generalized according to
\begin{align}
\label{eqn:comp_linear_system_standard_model_marangoni}
A_{\text{slip}}
\begin{pmatrix}
c_1 \\ c_2 \\ c_{35} \\ c_4
\end{pmatrix}
= \begin{pmatrix} 0 \\ -\frac{L}{V_0} \frac{\partial_{\tau_1} \sigma}{\eta} \\ 0 \\ 1\end{pmatrix}.
\end{align}
In this case, the solution reads as ($\theta \not\in \{0,\pi/2,\pi \}$)
\begin{align}
c_1 = c_4 = 0, \quad c_{35} = 1, \quad c_2 = -\frac{1}{2} \left( \cot 2\theta - \frac{L}{V_0} \frac{\partial_{\tau_1} \sigma}{\visc} \csc 2\theta \right).
\end{align}
Application of the kinematic evolution equation \eqref{eqn:comp_theta_dot} leads to
\begin{align}
\dot\theta = \frac{1}{2} \left(\frac{V_0}{L} - \frac{\partial_{\tau_1} \sigma}{\visc} \right),
\end{align}
which is the generalization of \eqref{eqn:comp_theta_dot_slip_model} to the case with a non-constant surface tension coefficient. Hence, a regular quasi-stationary solution satisfying $\dot\theta = 0$ may exist if a surface tension gradient 
\begin{align}
\frac{\partial \sigma}{\partial \tau_1} = \frac{\visc V_0}{L} = \lambda V_0
\end{align}
is present at the moving contact line (see also \cite{Fricke.2019}). Note that for an advancing contact line ($V_0 > 0$), this means that the surface tension is locally increased at the moving contact line. Remarkably, the same formula for the surface tension gradient has been found by Sibley et al.\ using asymptotic methods requiring a finite pressure at the moving contact line (see \cite{Sibley2015}, Appendix A). This shows that surface tension gradients at the contact line can indeed regularize the singularity.
 
\section[Compatibility analysis for the model by Lukyanov and Pryer]{Compatibility analysis for the model\\ by Lukyanov and Pryer} 
\subsection{The model}
The model introduced in \cite{Lukyanov.2017} is based on the Interface Formation Model (IFM) due to Y. Shikhmurzaev. The main simplifying assumptions are that the flow is quasi-stationary (i.e.\ the interface is fixed in a co-moving frame) and can be described by the stationary Stokes problem
\begin{align}
\label{eqn:comp_stationary_stokes}
\divergence{v} = 0, \quad \nabla p = \visc \Delta v \quad \text{in} \ \Omega. 
\end{align}
The interfacial mass densities are assumed to be constant, i.e.
\[ \rho_s^{(i)} \equiv \text{const}. \]
This allows to eliminate the interface velocity $\vsigma$ from the model leading to a modification of the boundary conditions for the Stokes equations (see \cite{Lukyanov.2017} for a derivation). In particular, the presence of interfacial mass allows for a non-zero normal component of the fluid velocity at the solid wall. In the following, we abbreviate this model with the term ``CSM model'' (constant surface mass model). Note that, in contrast to the IFM, the CSM model does not assume the presence of surface tension gradients at the contact line. Instead, the dynamic contact angle is an \emph{input} parameter for the model.\\
\\
The resulting boundary conditions for \eqref{eqn:comp_stationary_stokes} are given by
\begin{align}
\inproduct{v}{n_1} = \alpha_1 \nabla_{\Gamma_1}\cdot{(\mathcal{P}_1 v)} \quad &\text{on} \ \Gamma_1,\label{eqn:comp_bc_1}\\
\inproduct{\tau_1}{D n_1} = 0 \quad &\text{on} \ \Gamma_1,\label{eqn:comp_bc_2}\\
\inproduct{v}{n_2} = \alpha_2 \nabla_{\Gamma_2}\cdot{(\mathcal{P}_2 v)} \quad &\text{on} \ \Gamma_2,\label{eqn:comp_bc_3}\\
2L\, \inproduct{\tau_2}{D n_2} = V_0 - \inproduct{\tau_2}{v} \quad &\text{on} \ \Gamma_2,\label{eqn:comp_bc_4}\\
\alpha_1 \inproduct{v}{\tau_1} = \alpha_2 (V_0 + \inproduct{v}{\tau_2}) \quad &\text{at} \ \Gamma_1 \cap \Gamma_2\label{eqn:comp_bc_5},\\
-p + 2\visc \, \inproduct{n_1}{D n_1} = \sigma \kappa_1 \quad &\text{on} \ \Gamma_1,\label{eqn:comp_bc_6}
\end{align}
where $\mathcal{P}_i = \mathds{1} - n_i \otimes n_i$ is an orthogonal projection operator, $L$ is the slip length, $V_0$ is the velocity of the solid wall in the co-moving reference frame of the contact line and
\[ \alpha_1 := \frac{\rho_s^{(1)}}{\rho}, \quad \alpha_2 := \frac{\rho_s^{(2)}}{2\rho}.  \]
Note that the ratios $\alpha_i$ have the dimension of a length (since the liquid bulk density $\rho$ has units of mass per volume). Therefore, it is convenient to introduce the dimensionless quantities
\[ \hat{\alpha}_i := \frac{\alpha_i}{L}. \]
Note that the normal stress condition \eqref{eqn:comp_bc_6} is formulated for a constant ambient pressure $p_{\text{ext}} = 0$.

\subsection{Compatibility analysis for regular solutions}
We consider again the linear expansion of the velocity field at the contact line given by \eqref{eqn:comp_linear_field}. The set of boundary conditions \eqref{eqn:comp_bc_1}-\eqref{eqn:comp_bc_6} leads to a system of algebraic equations that is discussed below.

\paragraph{Mass balance at the interfaces:} Note that the mass balance equations \eqref{eqn:comp_bc_1} and \eqref{eqn:comp_bc_3} involving the surface divergence operator $\nabla_{\Gamma_i} \cdot$ require some clarification. A short computation shows the relation
\begin{align*}
\nabla_{\Gamma_i} \cdot (\mathcal{P}_i v) &= \nabla_{\Gamma_i} \cdot (v - \inproduct{n_i}{v}n_i) \\
&= \nabla_{\Gamma_i} \cdot v - \inproduct{n_i}{v} \nabla_{\Gamma_i} \cdot n - \underbrace{\inproduct{n_i}{\nabla_{\Gamma_i}(\inproduct{n_i}{v})}}_{= \, 0} = \nabla_{\Gamma_i} \cdot v + \inproduct{v}{n_i} \kappa_i.
\end{align*}
Making use of the incompressibility condition $0 = \nabla \cdot v = \nabla_{\Gamma_i} \cdot v + \inproduct{n_i}{(\nabla v)n_i}$, we conclude that
\begin{align*}
\nabla_{\Gamma_i} \cdot (\mathcal{P}_i v) &= - \inproduct{n_i}{(\nabla v)n_i} + \inproduct{v}{n_i} \kappa_i = - \inproduct{n_i}{D\,n_i} + \inproduct{v}{n_i} \kappa_i.
\end{align*}
It follows that the mass balance equations \eqref{eqn:comp_bc_1} and \eqref{eqn:comp_bc_3} can be expressed as
\begin{equation}
\label{eqn:comp_surface_divergence_term}
\begin{aligned}
(1-\hat\alpha_i \hat\kappa_i) \inproduct{v}{n_i} + \alpha_i \inproduct{n_i}{D\, n_i} = 0,
\end{aligned}
\end{equation}
where $\hat\kappa_i = \kappa_i L$ is the dimensionless curvature. We assume the solid boundary to be flat, i.e.\ $\hat\kappa_2 = 0$.\\
\\
Note that \eqref{eqn:comp_surface_divergence_term} implies that the system of equations \eqref{eqn:comp_bc_1}-\eqref{eqn:comp_bc_5} does only involve the symmetric part of $\nabla v$, i.e.\ the rate-of-strain tensor $D$. It seems that one has only $4$ unknowns ($c_1,c_2,c_{35},c_4$) for the $5$ equations \eqref{eqn:comp_bc_1}-\eqref{eqn:comp_bc_5}. But note that the curvature of the free-surface $\kappa_1$ at the contact line is introduced as an additional unknown parameter in \eqref{eqn:comp_surface_divergence_term}.\\
\\
Evaluating the mass balance equation \eqref{eqn:comp_surface_divergence_term} for the free surface at the contact line yields
\begin{align*}
(1-\hat\alpha_1 \hat\kappa_1) \, \dot{m}_1 = \frac{\hat\alpha_1}{2} (2 c_2 \cos 2\theta + c_{35} \sin 2\theta),
\end{align*}
where $\dot{m}_1 := v \cdot n_1(0,0) = - c_1 \sin \theta + c_4 \cos \theta $ is the (dimensionless) mass flux to the free surface phase at the contact line. The mass balance equation for the solid surface at the contact line reads as (since $\hat\kappa_2 = 0$)
\[ \dot{m}_2 = \hat\alpha_2 c_2,  \]
where $\dot{m}_2 := v \cdot n_2(0,0) = -c_4$ is the dimensionless mass flux to the solid-liquid surface phase at the contact line. Note that $\dot{m}_i = 0$ vanishes if the interface $\Gamma_i$ does not carry mass, i.e.\ if $\hat\alpha_i = 0$.

\paragraph{Mass balance at the contact line:} Equation \eqref{eqn:comp_bc_5} expresses the fact that the contact line itself cannot store mass. In terms of the unknown coefficients it translates to
\[ \left(\hat\alpha_2 \cos \theta + \hat\alpha_1 \right) c_1 + \hat\alpha_1 c_4 \sin \theta  = -\hat\alpha_2. \]
Note that the latter equation is trivially satisfied for $\hat\alpha_1 = \hat\alpha_2 = 0$. Moreover, if the free surface does not carry mass it follows that
\begin{align}
\hat\alpha_2 c_1 \cos\theta  = -\hat\alpha_2.
\end{align}
Hence no regular solution exists for $\hat\alpha_1 = 0, \, \hat\alpha_2 \neq 0$ and $\theta = \pi/2$.

\paragraph{Zero stress condition and Navier Slip condition:} Both the condition on the tangential stress \eqref{eqn:comp_zero_stress_condition_coefficients} and the Navier slip condition \eqref{eqn:comp_navier_slip_standard_model} remain unchanged with respect to the standard model. Note, however, that in this case $c_1$ may be non-zero since the free surface is not a material interface.
\\

\begin{mdframed} \textbf{Summary:} A regular solution to the model \eqref{eqn:comp_stationary_stokes}-\eqref{eqn:comp_bc_5} satisfies the following equations for the unknown coefficients  $c_1,c_2,c_{35},c_4,\hat\kappa_1$:
\begin{align}
(1-\hat\alpha_1 \hat\kappa_1) (-c_1 \sin \theta + c_4 \cos \theta) - \frac{\hat\alpha_1}{2}(2 c_2 \cos 2\theta+ c_{35} \sin 2\theta) &= 0,\label{eqn:comp_system_1}\\
-2 c_2 \sin 2\theta + c_{35} \cos 2\theta &= 0,\label{eqn:comp_system_2}\\
\hat\alpha_2 c_2 + c_4 &= 0,\label{eqn:comp_system_3}\\
c_1 - c_{35} &=1,\label{eqn:comp_system_4}\\
\left(\hat\alpha_1 \, \cos \theta + \hat\alpha_2 \right) c_1 + \hat\alpha_1 c_4  \sin \theta &= -\hat\alpha_2,\label{eqn:comp_system_5}
\end{align}
where $\hat\alpha_1, \hat\alpha_2$ and $\theta$ are given (dimensionless) data.
\end{mdframed}
Note that equation \eqref{eqn:comp_system_1}, expressing the mass balance in the free surface phase, is non-linear, while equations \eqref{eqn:comp_system_2}-\eqref{eqn:comp_system_5} constitute a linear subsystem
\[ \ALukyanov \begin{pmatrix} c_1\\ c_2\\ c_{35}\\ c_4 \end{pmatrix} = \begin{pmatrix} 0\\ 0\\ 1\\ \boldsymbol{-\hat\alpha_2} \end{pmatrix} \]
with the matrix $\ALukyanov$ given by
\begin{align}
\ALukyanov = 
\begin{pmatrix}
0 & -2 \sin(2\theta) & \cos(2\theta) & 0\\
0 & \boldsymbol{\hat\alpha_2} & 0 & 1\\
1 & 0 & -1 & 0\\
\boldsymbol{\hat\alpha_1 \, \cos\theta +\hat\alpha_2} & 0 & 0 & \boldsymbol{\hat\alpha_1 \, \sin\theta}
\end{pmatrix}.
\end{align}
Note also that, in the limit of vanishing surface mass ($\hat\alpha_1, \, \hat\alpha_2 \rightarrow 0$), the nonlinear equation \eqref{eqn:comp_system_1} becomes linear and the set of equations \eqref{eqn:comp_system_1}-\eqref{eqn:comp_system_4} reduces to \eqref{eqn:comp_linear_system_standard_model} while \eqref{eqn:comp_system_5} becomes obsolete. In this sense, the CSM model is a generalization of the standard slip model discussed in Section~\ref{sec:standard_model}.

\subsection{Solution of the non-linear system}
The determinant of the system matrix $\ALukyanov$ is given as
\begin{align}
\det \ALukyanov = 2 \left(\hat\alpha_1 \cos\theta + \hat\alpha_2 \right) \sin 2\theta - \hat\alpha_1 \hat\alpha_2 \sin \theta \cos 2\theta.
\end{align}
In the special case $\theta=\frac{\pi}{2}$ this simplifies to $\det \ALukyanov = \hat\alpha_1\hat\alpha_2$. Clearly, the linear part of the problem is uniquely solvable in case $\det \ALukyanov\neq0$. If, moreover, the solution satisfies
\[ \dot{m}_1 = c_4 \cos \theta - c_1 \sin \theta \neq 0, \]
we obtain the dimensionless curvature $\hat\kappa_1 = \kappa_1 L$ of the free surface at the contact line from the relation \eqref{eqn:comp_system_1}. The curvature is not determined by the compatibility conditions if there is no mass flux in the surface phase $\Gamma_1$, i.e. if $\dot{m}_1 = 0$ (see Section~\ref{sec:standard_model}).

\paragraph{General solution:}
Provided that $\det \ALukyanov \neq 0$, one can uniquely solve the system of equations \eqref{eqn:comp_system_2} - \eqref{eqn:comp_system_5}. In fact, the general solution is given by the expression
\begin{align}
\label{eqn:comp_general_solution_linear_system}
\begin{pmatrix} c_1\\ c_2\\ c_{35}\\ c_4 \end{pmatrix} = \frac{1}{\det \ALukyanov} \begin{pmatrix} -(4\cos\theta + \hat\alpha_1 \cos 2\theta) \hat\alpha_2 \sin \theta \\ -(2\hat\alpha_2 + \hat\alpha_1 \cos \theta) \cos 2\theta\\ -2(2\hat\alpha_2+\hat\alpha_1 \cos\theta) \sin 2\theta\\ (2\hat\alpha_2 + \hat\alpha_1 \cos \theta) \hat\alpha_2 \cos 2\theta \end{pmatrix}.
\end{align}

\paragraph{Remarks:} 
\begin{enumerate}[(i)]
 \item It is easy to show that the condition
\begin{align}
\label{eqn:comp_invertibility_condition}
0 < \hat\alpha_2 < \frac{2 \hat\alpha_1}{2+\hat\alpha_1} 
\end{align}
is sufficient for $\det(\ALukyanov)  \neq 0$ on $(0,\pi)$ for given $\hat\alpha_1 > 0$.
 \item A short calculation using \eqref{eqn:comp_general_solution_linear_system} shows that equation \eqref{eqn:comp_system_1} can be simplified according to
 \begin{align}
 \label{eqn:comp_curvature}
 (1-\hat\alpha_1 \hat\kappa_1) \, \dot{m}_1 = -\frac{\hat\alpha_1}{\det\ALukyanov} (2\hat\alpha_2 + \hat\alpha_1 \cos \theta).
 \end{align}
 The latter equation is central for the regularity of solutions to the CSM model as we will discuss below.
 \item Provided that $0 < \hat\alpha_2 \leq \frac{\hat\alpha_1}{2}$, there is a unique contact angle given by
 \[ \thetastar = \arccos\left(-\frac{2 \hat\alpha_2}{\hat\alpha_1}\right) > \frac{\pi}{2} \]
 which makes the right-hand side of equation~\eqref{eqn:comp_curvature} equal to zero. In this case, it follows that 
 \[ \hat\kappa_1(\thetastar) = \frac{1}{\hat\alpha_1} \quad \text{if} \quad \dot{m}_1(\thetastar) \neq 0.\] 
 Otherwise (i.e.\ for $\theta = \thetastar$ and $\dot{m}_1(\thetastar) = 0$), the above equation \eqref{eqn:comp_curvature} becomes obsolete. In this case the boundary conditions are compatible but the curvature at the contact line is \emph{not} determined by the compatibility conditions.
 \item On the other hand, if there is a set of parameters $\{ \theta_s \neq \thetastar, \hat\alpha_1, \hat\alpha_2\}$ such that $\dot{m}_1(\theta_s,\hat\alpha_1,\hat\alpha_2) = 0$, then there is \emph{no} regular solution to the model for this choice of parameters. Instead, the curvature $\hat\kappa_1$ becomes singular as $\theta \rightarrow \theta_s$ (see Section~\ref{sec:examples}).
\end{enumerate}

\subsection{Singularities in the model by Lukyanov and Pryer}\label{sec:subsection_singular_points}
\begin{figure}
 \centering
 \includegraphics[width=9cm]{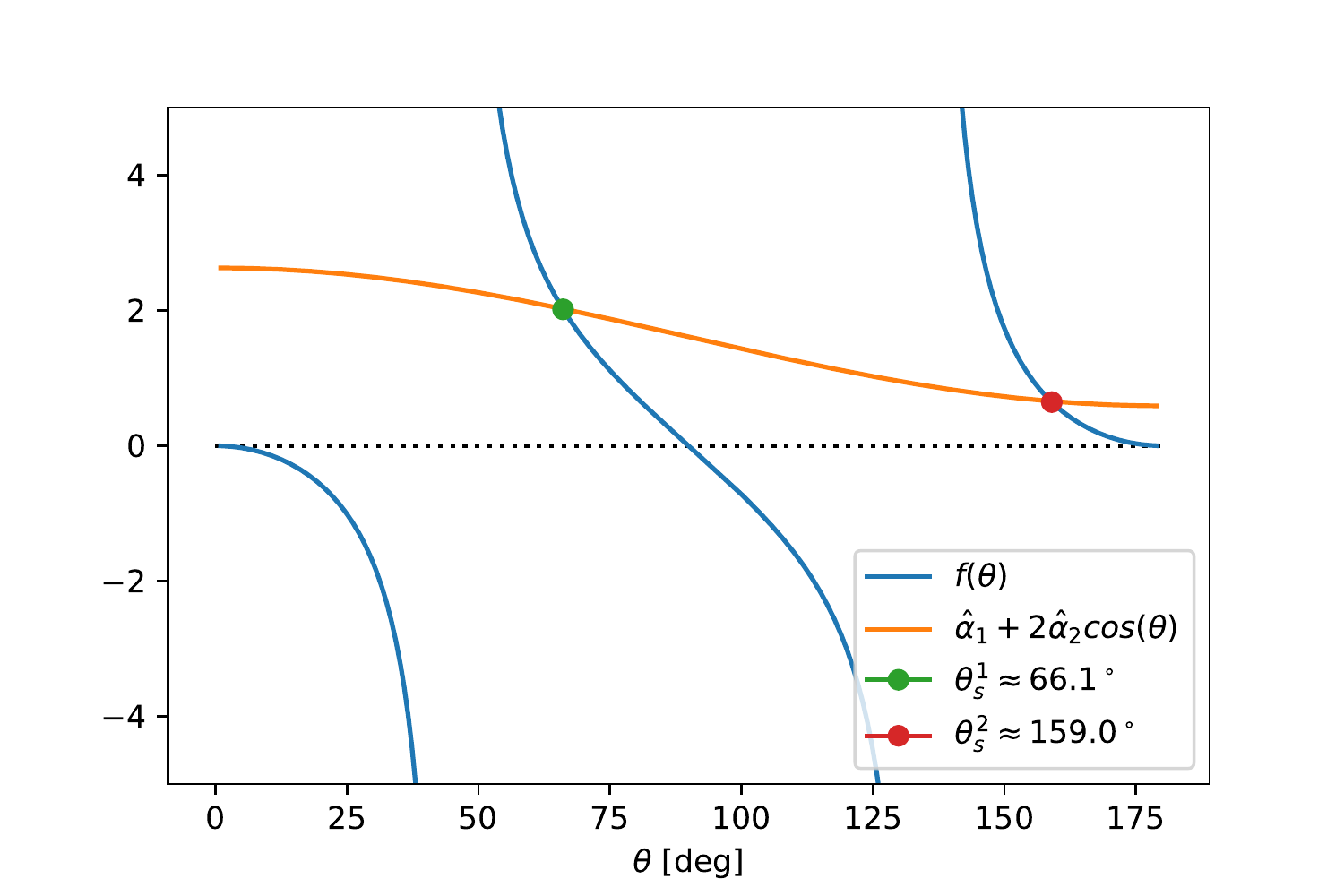}
 \caption{Roots of the mass flux $\dot{m}_1$ for $\hat\alpha_1 = 1.6$ and $\hat\alpha_2 = 0.51$.}
 \label{fig:root_of_mass_flux}
\end{figure}

We will now show that for any choice of surface mass densities $\hat\alpha_1, \, \hat\alpha_2 > 0$ satisfying the invertibility condition \eqref{eqn:comp_invertibility_condition}, there is always at least one choice of the contact angle $\theta_s < \frac{\pi}{2}$ such that no regular solution exists for the parameters $\{\hat\alpha_1,\hat\alpha_2, \theta_s \}$.\\
\\
The mass flux $\dot{m}_1$ can be computed from the general solution \eqref{eqn:comp_general_solution_linear_system} leading to the formula
\[ \dot{m}_1 = - c_1 \sin \theta + c_4 \cos \theta = \frac{\hat\alpha_2}{\det \ALukyanov} \left(4 \cos \theta \sin^2 \theta + [\hat\alpha_1 + 2 \cos \theta \, \hat\alpha_2 ] \cos 2\theta \right). \]
Hence the roots of $\dot{m}_1$ are the solutions of
\begin{align}
\label{eqn:comp_roots_dotm1}
\hat\alpha_1 + 2 \hat\alpha_2 \cos\theta  = f(\theta), 
\end{align}
where $f(\theta) = - (4 \cos \theta \sin^2 \theta)/\cos 2\theta$. Since the left-hand side of \eqref{eqn:comp_roots_dotm1} is monotonically decreasing with $\hat\alpha_1 + 2\hat\alpha_2 \cos(\frac{\pi}{2}) = \hat\alpha_1 > 0$, there is always a solution $\theta_s^1 < \frac{\pi}{2}$ of \eqref{eqn:comp_roots_dotm1}. Since the right-hand side of \eqref{eqn:comp_curvature} is non-zero on $[0,\frac{\pi}{2}]$, it follows that \eqref{eqn:comp_curvature} has no solution for $\{\hat\alpha_1,\hat\alpha_2, \theta_s^1 \}$ and no regular solution exists for the latter set of parameters.\\
Figure~\ref{fig:root_of_mass_flux} shows the example $\hat\alpha_1 = 1.6$ and $\hat\alpha_2 = 0.51$ given in \cite{Lukyanov.2017}. In this case we have a singular point at
\[ \theta_s^1 \approx 66.1^\circ \]
and a second root at $\theta_s^2 \approx 159^\circ$. Since in this case $\thetastar \neq \theta_s^2$, it follows that the second root of $\dot{m}_1$ also corresponds to a singularity of the model (see Section~\ref{sec:examples}).

\subsection{Pressure at the moving contact line}
We can now evaluate the normal stress condition \eqref{eqn:comp_bc_6} at the free surface. If the curvature $\kappa_1(0,0)$ is uniquely determined by the compatibility conditions, we also obtain the pressure at the contact point. The non-dimensional form of equation \eqref{eqn:comp_bc_6} reads
\begin{align}
\hat{p} = -\frac{\kappa_1 L}{\ca} + \frac{\inproduct{n_1}{2Dn_1}}{V_0/L}.
\end{align}
Here we defined the non-dimensional quantities
\[ \ca := \frac{\visc V_0}{\sigma}, \quad \hat{p} := \frac{p}{(\visc V_0)/L} = \frac{L}{\sigma} \, \frac{p}{\ca}. \]
At the contact point, we have
\begin{align*}
\frac{\inproduct{n_1}{2Dn_1}}{V_0/L}(0,0) = - 2 c_2 \cos 2\theta   -2 c_{35} \sin 2\theta.
\end{align*}
Therefore, the dimensionless pressure at the contact point is given by
\begin{align}
\label{eqn:comp_pressure_formula}
\boxed{\hat{p} = -\frac{\hat\kappa_1}{\ca} - 2 (c_2 \cos 2\theta + c_{35} \sin 2\theta).}
\end{align}
Hence the relation for the pressure (relative to the ambient pressure) reads as
\[ p = -\kappa_1 \sigma - \frac{2 \visc V_0}{L} (c_2 \cos 2\theta + c_{35} \sin 2\theta). \]
In particular, the pressure converges to the Laplace pressure $p_{\text{L}} = - \kappa_1 \sigma$ as $V_0 \rightarrow 0$.

\subsection[Special case]{Special case $\theta=\pi/2$}
The system of equations \eqref{eqn:comp_system_1}-\eqref{eqn:comp_system_5} is substantially simplified in the case $\theta = \pi/2$. In this case, the nonlinear equation \eqref{eqn:comp_system_1} reads as
\begin{align}
\label{eqn:comp_nonlinear_equation_90_degrees}
-(1 - \hat\alpha_1 \hat\kappa_1) c_1 + \hat\alpha_1 c_2 = 0
\end{align}
and the matrix $\ALukyanov$ is given by
\begin{align*}
\ALukyanov(\theta=\pi/2) = 
\begin{pmatrix}
0 & 0 & -1 & 0\\
0 & \hat\alpha_2 & 0 & 1\\
1 & 0 & -1 & 0\\
\hat\alpha_2 & 0 & 0 & \hat\alpha_1
\end{pmatrix}.
\end{align*}
The solution of the linear system is given by
\begin{align}
\label{eqn:comp_solution_csm_90_degrees}
\begin{pmatrix} c_1\\ c_2\\ c_{35}\\ c_4 \end{pmatrix} = \begin{pmatrix} 1 \\ 2/\hat{\alpha}_1 \\ 0 \\ -2\hat\alpha_2/\hat{\alpha}_1 \end{pmatrix}.
\end{align}
According to \eqref{eqn:comp_nonlinear_equation_90_degrees}, it follows that the curvature at the contact line is
\begin{align}
\label{eqn:comp_cl_curvature_90_degrees}
\boxed{\kappa_1 L = -\frac{1}{\hat\alpha_1} \leq 0.}
\end{align}
So the curvature is always negative, i.e.\ the free surface is always convex locally at the contact line for regular solutions with $\theta = \pi/2$. Moreover, the curvature becomes singular as $\hat\alpha_1 \rightarrow 0$, which can be understood as the transition to the standard model. Interestingly, both the curvature and pressure at the contact line do not depend on the surface mass density $\hat\alpha_2$ in the liquid-solid phase. Note, however, that according to \eqref{eqn:comp_solution_csm_90_degrees} the coefficient $c_4$ in the expansion \eqref{eqn:comp_linear_field} still depends on $\hat\alpha_2$.\\
\\
For the dimensionless pressure at the contact line we obtain, according to \eqref{eqn:comp_pressure_formula}, the relation
\begin{align}
\boxed{\hat{p} = \frac{1}{\hat\alpha_1} \left( 4 + \frac{1}{\text{Ca}} \right).}
\end{align}
Therefore, the pressure (relative to the gas phase) is zero for $\theta = \frac{\pi}{2}$ and $\ca = -\frac14$, independently of $\hat\alpha_1$ and $\hat\alpha_2$. For the pressure in physical units we find
\begin{align*}
p = \frac{\sigma \ca}{L} \hat{p} = \frac{\sigma}{L \hat\alpha_1} \left(1 + 4\ca \right).
\end{align*}
Making use of the expression \eqref{eqn:comp_cl_curvature_90_degrees} for the mean curvature, we conclude
\begin{align*}
p = - \sigma \kappa_1(1 + 4\ca).
\end{align*}
 
\section{Comparison with results by Lukyanov and Pryer}
\label{sec:examples}
\begin{figure}[ht]
 \centering
 \includegraphics[width=9cm]{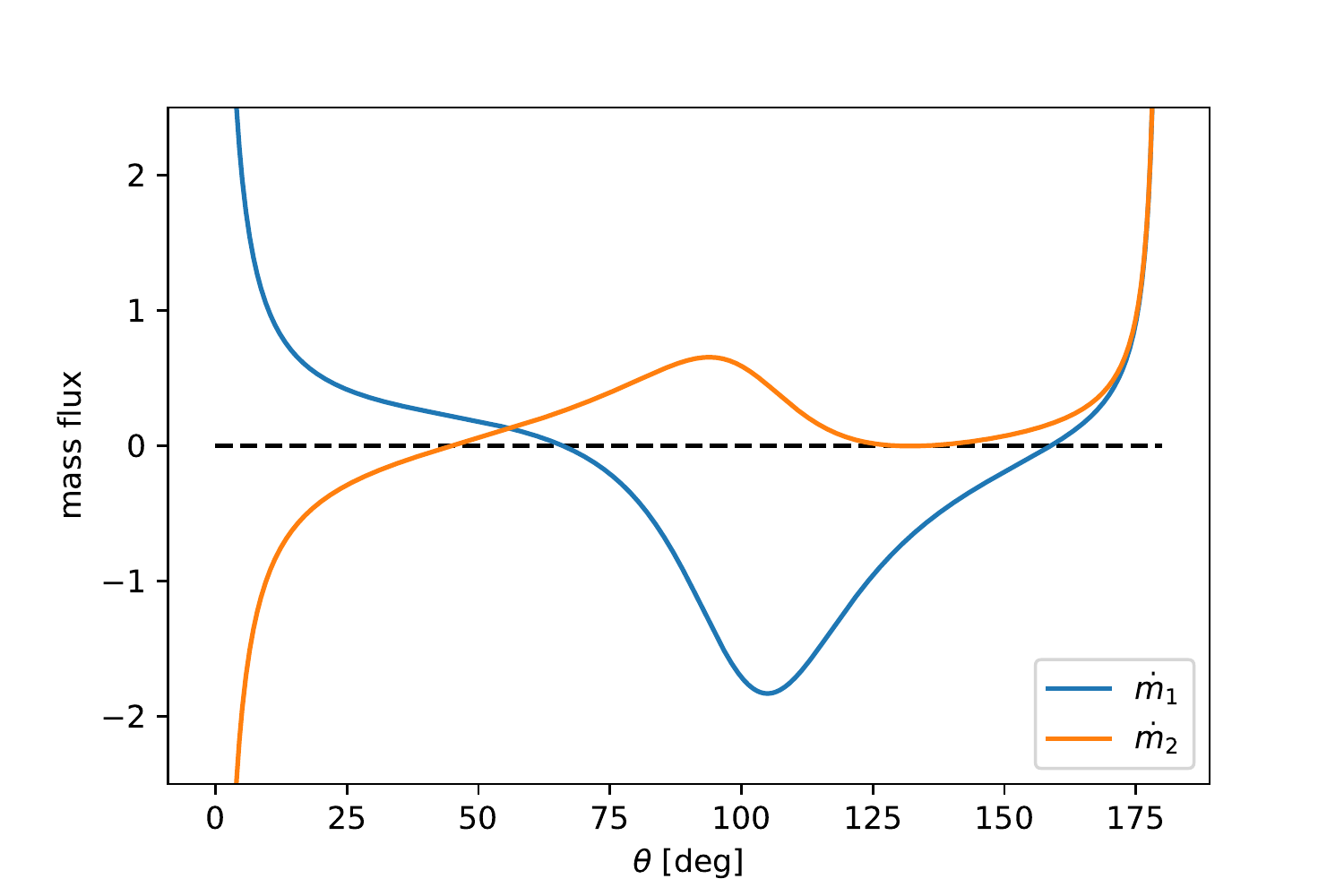}
 \caption{Non-dimensional mass fluxes for $\hat\alpha_1 = 1.6$ and $\hat\alpha_2 = 0.51$.}
 \label{fig:mass_flux}
\end{figure}

\begin{figure}[ht]
 \subfigure[Curvature.]{\includegraphics[width=7cm]{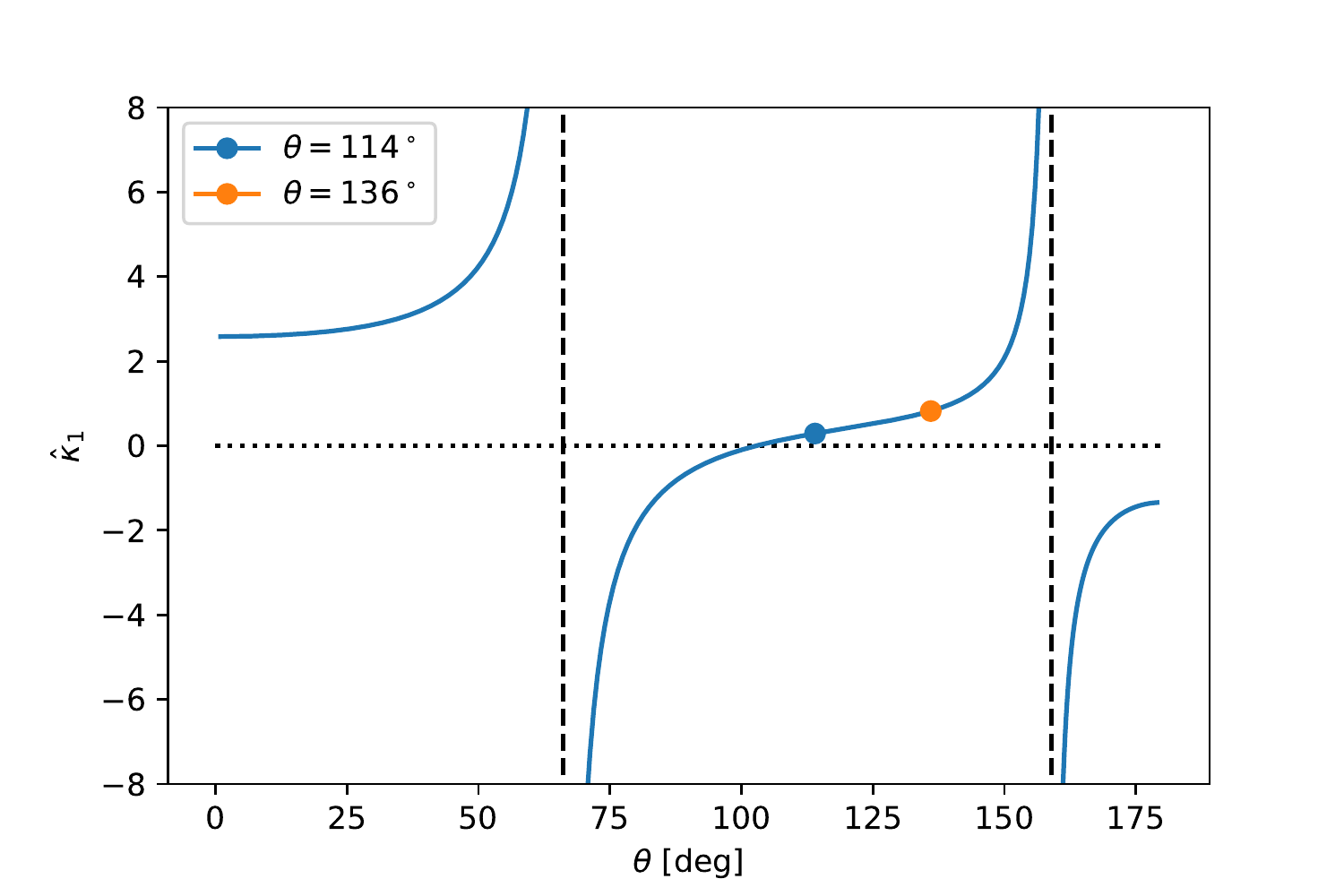}\label{fig:curvature}}
 \subfigure[Pressure ($\ca = 0.057$).]{\includegraphics[width=7cm]{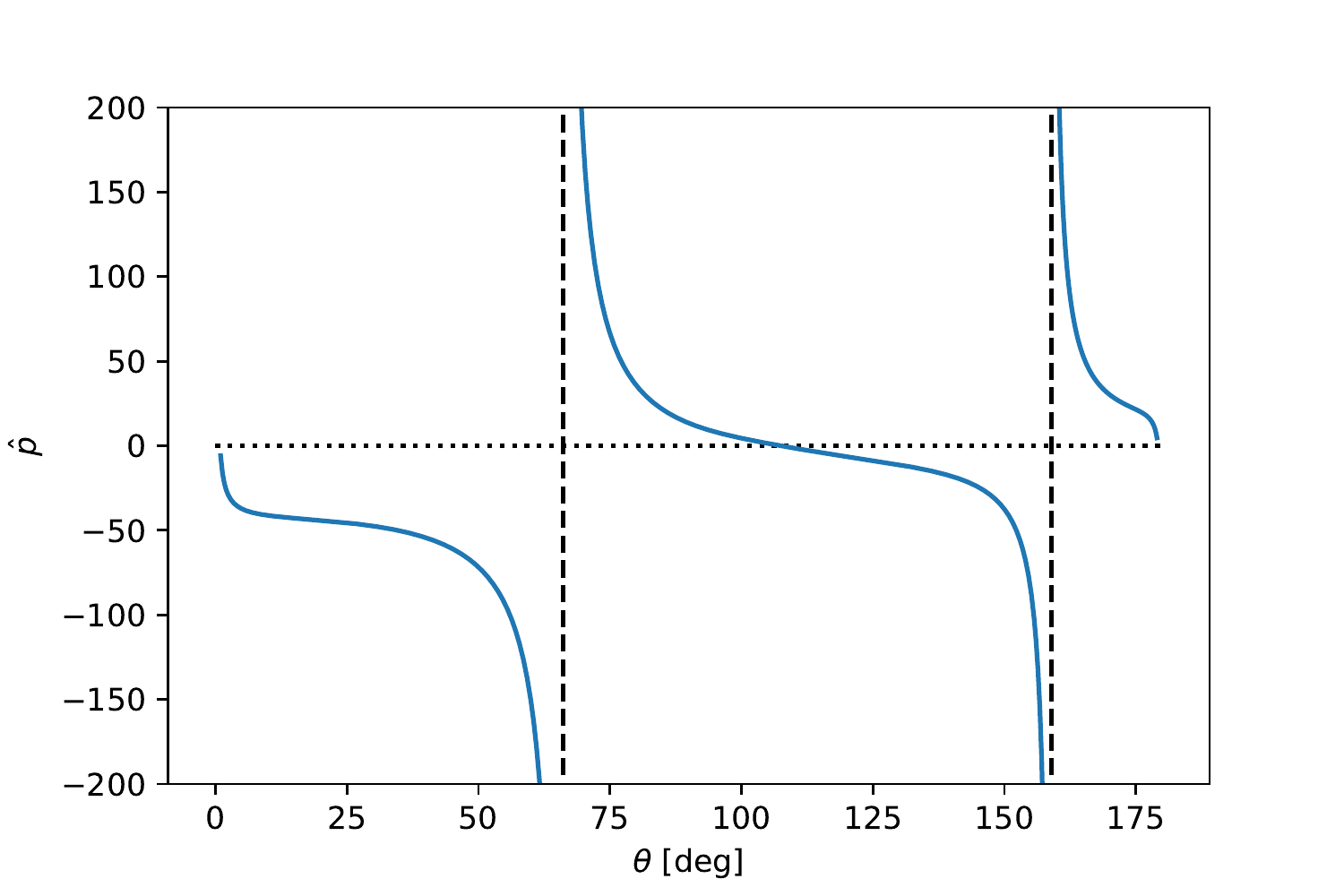}\label{fig:pressure}}
 \caption{Dimensionless curvature and pressure at the contact line for $\hat\alpha_1 = 1.6$ and $\hat\alpha_2 = 0.51$.}
\end{figure}

In the following, we revisit some examples given in \cite{Lukyanov.2017}. The reported values for the interfacial mass densities are
\[ \hat\alpha_1 = 1.6, \quad \hat\alpha_2 = 0.51. \]
The latter values are obtained from Molecular Dynamics (MD) simulations (see \cite{Lukyanov.2017} for details). The matrix is invertible for all $\theta \in (0,\pi)$ since \eqref{eqn:comp_invertibility_condition} is satisfied. Therefore, the non-dimensional mass fluxes $\dot{m}_i$ at the contact line can be computed from the general solution \eqref{eqn:comp_general_solution_linear_system} (see Figure~\ref{fig:mass_flux}). It is observed that both $\dot{m}_1$ and $\dot{m}_2$ become singular for $\theta \rightarrow 0, \pi$. Moreover, the mass flux $\dot{m}_1$ has two roots, corresponding to (see Section~\ref{sec:subsection_singular_points})
\[ \theta_s^1 \approx 66.1^\circ \quad \text{and} \quad \theta_s^2 \approx 159.0^\circ. \]
The roots of the mass flux $\dot{m}_1$ lead to singularities in the curvature according to \eqref{eqn:comp_curvature}, see Figure~\ref{fig:curvature}. Consequently, the singularities are also present in the pressure at the contact line (see Figure~\ref{fig:pressure} for a plot with fixed $\ca$). In the present example, the curvature has a root at $\theta_0 \approx 102.9^\circ$.  The curvature is positive for contact angles in between $\theta_0$ and $\theta_s^2$, i.e.\ the interface is locally \emph{concave}.\\
\\
The values for the dimensionless curvature and pressure at the contact line for the examples given in \cite{Lukyanov.2017} are summarized in Table~\ref{tab:results}. For the cases (i)-(ii) and (iv), the sign of the curvature does not agree with the macroscopic form of the interface as reported in \cite{Lukyanov.2017}. Hence, the present mathematical analysis predicts a bending of the interface close to the contact line even for the nanodroplet considered in \cite{Lukyanov.2017}. The absolute value of the curvature for the case (iii) is much larger than the macroscopic curvature of the interface reported in \cite{Lukyanov.2017}. In fact, the contact angle $\theta = 65^\circ$ is close to the singular point $\theta_s^1 \approx 66.1^\circ$. This is also the reason for the extremely low dimensionless pressure of $\hat{p} \approx - 711$ (measured relative to the ambient pressure). This value is about three orders of magnitude lower than the pressure reported in \cite{Lukyanov.2017} (being approximately $-0.9$).

\begin{table}[ht]
	\centering
	\begin{tabular}{ c| c| c| c| c| c| c|}
        Case & $\hat\alpha_1$ & $\hat\alpha_2$ & $\theta$ & $\ca$ & $\hat\kappa_1$ & $\hat{p}$\\
        \hline
        (i) & 1.6 & 0.51 & $136^\circ$ & 1.14 & 0.81 & -1.13 \\
        \hline
        (ii) & 1.6 & 0.51 & $114^\circ$ & 0.34 & 0.29 & 0.77 \\
        \hline
        (iii) & 1.6 & 0.51 & $65^\circ$ & 0.057 & 40.70 & -711.64 \\
        \hline
        (iv) & 0.62 & 0.33 & $123^\circ$ & 0.69 & 1.02 & 12.64
	\end{tabular}
	\caption{Dimensionless pressure and curvature for the examples given in \cite{Lukyanov.2017}.}
	\label{tab:results}
\end{table} 
\section{Conclusion}
The method of compatibility analysis is applied to two continuum mechanical models of dynamic wetting, namely the standard Navier slip model and the model introduced by Lukyanov and Pryer \cite{Lukyanov.2017}. It is shown that no quasi-stationary regular solutions with a moving contact line exist for the standard model in the absence of surface tension gradients. Even if the contact angle is allowed to vary, we showed that regular solutions behave unphysical (see also \cite{Fricke.2019}). Therefore, physical solutions of the standard model with constant surface tension must be singular at the moving contact line. Moreover, it is shown that a surface tension gradient at the moving contact line may give rise to regular solutions (see also \cite{Sibley2015},\cite{Fricke.2019}).\\
\\
The model \cite{Lukyanov.2017} allows to store mass on the liquid-gas and liquid-solid interfaces (in the general framework of \cite{Shikhmurzaev.1993}). However, the surface mass density is assumed to be constant in space and time leading to modified boundary conditions. It is shown that the standard model is recovered in the limit of vanishing surface mass densities (i.e.\ $\hat\alpha_1, \, \hat\alpha_2 \rightarrow 0$). The compatibility analysis for the model shows that regular solutions with finite pressure and curvature at the contact line are possible. In fact, the curvature and the pressure at the contact line can be computed from the compatibility conditions provided that the mass flux $\dot{m}_1$ to the free surface at the contact line is non-zero. On the other hand, if for certain model parameters the mass flux $\dot{m}_1$ goes to zero, then the solution becomes singular very much like in the standard model (where $\dot{m}_1$ is always zero). It is shown that such a singular point exists for every pair of interfacial mass densities $(\hat\alpha_1,\hat\alpha_2)$ satisfying \eqref{eqn:comp_invertibility_condition}. Hence the model does not always ``cure'' the singularity completely. Explicit expressions for the pressure and the curvature at the contact line are derived for the special case of $\theta = \pi/2$. Interestingly, the latter values for $\theta = \pi/2$ do not depend on the surface mass density $\hat\alpha_2$ of the liquid-solid interface.\\
\\
The numerical values for the local curvature and the pressure at the contact line are compared with the continuum mechanical simulations of the model reported in \cite{Lukyanov.2017} showing a quantitative and even qualitative disagreement. A possible explanation is a strong bending of the interface close to the contact line which is not resolved by the simulations in \cite{Lukyanov.2017}.\\
\\
We emphasize that the method of compatibility analysis discussed here does only provide information \emph{locally at} the contact line (or, in two dimensions at the contact point). There is no statement about the change in curvature or pressure in the vicinity of the contact line, which might explain the discrepancy to the results in \cite{Lukyanov.2017}. However, the mathematical method of compatibility analysis is quite general and, therefore, applicable to a variety of models, possibly allowing for new insights into the complex problem of dynamic wetting. 
\paragraph{Acknowledgments:} We kindly acknowledge the financial support by the German Research Foundation (DFG) within the Collaborative Research Centre 1194 “Interaction of Transport and Wetting Processes” – Project-ID 265191195 , subproject B01. The authors thank S.~Afkhami, T.~Gambaryan-Roisman and L.~Pismen for organizing the workshop “\textit{Challenges in nanoscale physics of wetting phenomena}” hosted by the Max Planck Institute for the Physics of Complex Systems, Dresden, August 2019.

\end{document}